\documentclass[pra,preprint,preprintnumbers,amsmath,amssymb,nofootinbib]{revtex4}

\usepackage{graphicx}
\usepackage{amssymb}
\usepackage{amsmath}
\usepackage{amsfonts}
\usepackage{mathrsfs}
\usepackage{color}
\usepackage{ulem}
\usepackage{physics}
\usepackage[english]{babel}
\usepackage[utf8]{inputenc}
\usepackage[T1]{fontenc}
\usepackage{mathptmx}
\usepackage{etoolbox}

\makeatletter
\def\@email#1#2{%
 \endgroup
 \patchcmd{\titleblock@produce}
  {\frontmatter@RRAPformat}
  {\frontmatter@RRAPformat{\produce@RRAP{*#1\href{mailto:#2}{#2}}}\frontmatter@RRAPformat}
  {}{}
}%
\makeatother

\begin{document}

\newcommand{\half}{\frac{1}{2}}
\newcommand{\pcl}[1]{#1_{\mathrm{p}}}

\bibliographystyle{prsty}

\title{A version of de Broglie's double solution theory reproducing Landau's quantization in a uniform magnetic field }

\author{Pierre Jamet  and Aur\'elien Drezet}
\affiliation{Institut N\'eel, UPR 2940, CNRS-Universit\'e Joseph Fourier, 25, rue des Martyrs, 38000 Grenoble, France}

\date{\today}

\begin{abstract}

\end{abstract}
 \maketitle
 \section*{Introduction}
    Quantization of various physical quantities like momentum or energy is the hallmark of modern quantum mechanics. In fact, it even gave it its name, due to the importance of this aspect in the development of the old quantum theory as developped originally by Planck, Einstein, Bohr and many others in the first decades of the \textsc{xx}$^{th}$ century. However, it is well known today -- and it was already at the time -- that quantization is actually a property of waves in a broad sense, rather than a specificity of the microscopic world. In that context the paradigmatic discovery was the wave particle duality associated with the name of albert Einstein for the photon and Louis  de Broglie for the electron and other particles~\cite{Broglie1}. The idea of de Broglie to associate a wave to a particle soon evolved into the family of so called pilot wave theories~\cite{Broglie2,Solvay}. More precisely, the goal of such models was to attach to each particle a wave, which would guide its dynamics so that some aspects of undulatory mechanics could then be applied to it. The idea was inspired by works made by Fermat and Hamilton showing some analogies between optics and classical mechanics of point-like objects.    For instance, if we consider the case of an electron orbiting around the nucleus of an atom, its associated wave would have to satisfy a periodicity condition along the perimeter of the trajectory, which would then translate to a quantization of its physical properties, e.g. wavelength, energy, frequency and so on. The electron, being so closely related to the wave, would then acquire the same restrictions on its own properties, and a quantization of its angular momentum, energy, radius and velocity would result. One way to achieve this duality between a wave and a particle was devised by de Broglie, who postulated the existence of an internal periodic phenomenon inside each particle, a clock of sorts, which would have to be synchronized with the oscillations of the wave. If, then, the frequency of the wave had to be quantized due to some boundary condition, the particle would also need to constrain its own dynamics in order to maintain the synchronization during the motion.
Remarkably, the two ideas first introduced by de Broglie almost a century ago as the basis of pilot wave theories, namely the existence of a phase wave and its synchronization with the oscillation of some property of the particle itself, still remain in some form or another in current approaches. 

In fact in the past two decades, macroscopic hydrodynamic experiments have been performed that manage to reproduce a good number of the properties of quantum objects~\cite{Couder,HQA,Fort}. In light of these recent developments, one can wonder how we could translate these results to a microscopic model, i.e. applied to electrons and atoms rather than oil droplets for example. Compared to other quantum theories and interpretations (e.g., the Copenhagen interpretation or the Many Worlds intepretation), this approach has some unique constraints, the main one being that this model needs to remain a mechanical one all throughout the development. It means that we need to have ``hidden'' variables and, of course, that the field and interactions must propagate locally at all times. On the other hand, it will be necessary to deviate from the usual macroscopic descriptions as we will see throughout this document. The model should also differs from the so called `Bohmian mechanics' \cite{Bohm} which is a logical developement of the original de Broglie pilot wave approach (realized after the rediscovery of the theory by Bohm in 1952). In Bohmian mechanics a wave guides a particle  but this is done in the  configuration space and not in the 3D physical space. Also the physical nature of this guiding wave is ambiguous and mysterious since it acts on the particle but there is no back-reaction on the wave. De Broglie didn't consider the pilot wave approach as a good starting point.  He hoped (somehow in agreement with Einstein) to find a mechanical explanation (the double solution \cite{Broglie3}) justifying the existence of the coupling between a localized object and an extended wave.\\     
\indent In the following, we will present one such theoretical mechanical model developed in recent years \cite{Jamet1,Jamet2,Jamet3} following de Broglie's intuition about wave-particle duality, and show that it is indeed able to predict the Landau quantization of an electron orbiting in a homogeneous magnetic field \cite{Landau}. This result is to be put in comparison with an experimental demonstration of the Landau levels, achieved in a system consisting of a droplet bouncing on a vibrating bath \cite{Fort}. Compared to macroscopic experiments however, some different ingredients are needed as mentioned. For instance, it is required to work under a relativistic framework, but also to make use of complex quantities -- in particular a complex field -- in order to take into account the electric charge. While forcing us to deviate from what we are used to in hydrodynamical descriptions, these different aspects have the benefit of giving a really robust and self-consistent theory, so that we can for example use a completely covariant formalism in an elegant way.

The model that is presented here is an extension of one that has been developped in the past three years (motivated by a previous work \cite{Borghesi}). As we will see, it makes use of the fact that if we couple a scalar field with a point-like particle, there exists a regime where the equations of motion for the various objects decouple from each other. In that regime, called transparency, the initially complex and chaotic dynamics can become linear and stable on top of being able to reproduce results of the quantum theory. This result is of course very promising as an attempt to give a completely mechanical description of quantum objects, and in particular as it relates to the hydrodynamics experiments mentioned earlier. It is also a satisfying illustration of wave-particle duality, since in that specific regime the two objects no longer affect eachother and are so closely linked that they behave as one, so that to distinguish them becomes extremely challenging. We will first see in Section I how the model itself is built and how it relates to de Broglie's original ideas on wave-particle duality as well as pilot wave theories. In Section II, we will obtain the expressions for the different quantities defining the dynamics of the particle. Finally in Section III, we will derive the complete analytical solutions for the field, and conclude by comparing these results with the well-known Landau effect.

\section{Description of the system}
Inspired by the ideas of de Broglie, and more specifically his double solution theory, our goal is to build a simple mechanical model \cite{Jamet1,Jamet2,Jamet3} able to reproduce some of the quantum properties of a particle, for example here an electron. We start off with some basic ingredients in order to construct a pilot wave model, namely a complex scalar field $u(x)=u(t,\mathbf{x})\in \mathbb{C}$  (here $x^\mu:=[t,\mathbf{x}]\in \mathbb{R}^4$ denotes the relativistic position four-vector) acting as the guiding wave which we couple to a point-like particle of mass $m_\mathrm{p}$ located at the position $\mathbf{x}_\mathrm{p}(t)\in \mathbb{R}^3$. This particle will be constrained to live on the field at all times, meaning that we introduce another degree of freedom $z_\mathrm{p}(t)\in \mathbb{C}$ which one can think of as an internal oscillation analogous to de Broglie's internal clock, later synchronized with the oscillations of the field. To account for this natural periodic phenomenon for the particle, one needs to add a term akin to a harmonic force acting on this quantity $z_\mathrm{p}$, with a proper frequency $\Omega_\mathrm{p}$, or equivalently a stiffness $m_\mathrm{p}\Omega_\mathrm{p}^2$.
Our system~\footnote{In the following we work with covariant notations, and in particular we have the contravariant space-time four-vector $x^\mu:=[t,\vb{x}]$, the four-potential $A^\mu := [0,\vb{A}]$ and four-gradient $\partial^\mu := [\partial_t, -\grad]$, in a metric $g_{\mu\nu}= {\mathrm{diag}}(1,-1,-1,-1)$.} is thus characterized by the action 
\begin{equation}
    \begin{split}
		I = &-\int\bqty{\pcl{m} - \half m_{\mathrm{p}}\sigma\pqty{\vqty{\dot{z}(\tau)}^2 - \pcl{\Omega}^2 \vqty{z(\tau)}^2}}\dd\tau - e\int A(\pcl{x}(\tau))\pcl{\dot{x}}(\tau)\dd\tau \\
		&+ \int \left\lbrace \mathcal{N}(\tau)\bqty{z(\tau) - u(\pcl{x}(\tau))}^* + \mathcal{N}^*(\tau)\bqty{z(\tau) - u(\pcl{x}(\tau))}\right\rbrace\dd\tau \\ 
		&+ T\int\bqty{(Du)(Du)^* - \omega_0^2 u u^*}\dd^4 x 
	\end{split}
	\label{eq:action}
\end{equation}
which is split into three components:
\begin{itemize}
    \item[\textit{i)}] a Lagrangian for the particle integrated over the proper time $\tau$ comprised of a usual relativistic kinetic term, the internal degree of freedom, as well as the local interaction with a four-potential $A_\mu(x)$ through a coupling constant $e=-|e|<0$ (e.g. an electric charge for the `electron').
    \item[\textit{ii)}] a lagrangian density integrated over space-time for the field $u$, 
    with the covariant derivative $D_\mu = \partial_\mu + i e A_\mu$ using the same four-potential as it appears for the particle.
    \item[\textit{iii)}] an interaction Lagrangian between the field and the particle integrated over the proper time, making use of a Lagrange multiplier $\mathcal{N}(\tau)$ interpreted as the reaction force of the field on the particle.
\end{itemize}
 We stress that we here use the proper time $\tau$ as a parameter to label the position of the particle $x(\tau)$ along its trajectory.
Minimizing this action $\delta I = 0$ soon leads to a system of coupled Euler-Lagrange differential equations, from which we can extract a regime called transparency where the force $\mathcal{N}(\tau)$ vanishes, meaning that the particle and the field are no longer affecting each other. 
This specific regime is an interesting example of a wave-particle duality, in the sense that the two objects behave as one and share most of their respective properties, as we will see after obtaining the detailed solutions.

In the case of the transparency, the equations decouple, apart from the holonomic constraint
\begin{equation}
    z_\mathrm{p}(\tau) - u(\tau,x_\mathrm{p}(\tau)) = 0.
    \label{eq:holonomic}
\end{equation}
The dynamics of $z_\mathrm{p}(\tau)$ then becomes a simple harmonic motion at the proper frequency $\Omega_\mathrm{p}$
\begin{equation}
    z_\mathrm{p}(\tau) = z_0 e^{-i\Omega_\mathrm{p}\tau},
\end{equation}
while the particle and the field follow the equations
\begin{align}
    m_\mathrm{eff.}\ddot{x}_{\mathrm{p}\mu}(\tau) &= e F_{\mu\nu}(x_\mathrm{p}(\tau))\dot{x}_\mathrm{p}^\nu\label{eq:motion_x}\\
    \pqty{D^2 + \omega_0^2}u(t,x) &= 0\label{eq:motion_u}
\end{align}
with $m_\mathrm{eff.} = m_\mathrm{p}\pqty{1 + \sigma \Omega_\mathrm{p}^2\abs*{z_0}^2}$ an effective mass that takes into account both the usual inertia $m_\mathrm{p}$ and the oscillatory motion along $z$, $F_{\mu\nu} = \partial_\mu A_\nu - \partial_\nu A_\mu$ the electromagnetic tensor, $D_\mu = \partial_\mu + i e A_\mu$ the covariant derivative and finally $\omega_0$ the mass of the scalar field. We chose to consider a massive scalar field to keep the model as general as possible. We can however also decide to use a massless one where $\omega_0 \to 0$ as we did in other works at a later time. The advantage of using a Klein-Gordon equation for $u$ is to be able to give the correct non-relativistic limit to the field as we will see at the end of Section III, which a simple d'Alembertian would not allow due to the nature of the problem at hand, and especially the fact that we consider a pure magnetic field as our external force.\\

\indent At this point we can introduce another of de Broglie's ideas in order to give a qualitative description of the solutions. Indeed, if we consider that the particle must see, in its proper frame $\mathcal{R}^\prime$, a stationary field $u^\prime(t^\prime,\varphi^\prime) = f(t^\prime)g(\varphi^\prime)$, it follows quickly that along the path followed by the particle, the field must be written as a sum of two counterpropagating plane waves
\begin{equation}
    u(t,\varphi) = u_+ + u_- = \frac{u_0}{2}\pqty{e^{i(\omega_+ t - k_+ \rho_\mathrm{p}\varphi)} + e^{i(\omega_- t - k_- \rho_\mathrm{p}\varphi)}},
\end{equation}
or alternatively, a product of a phase and group wave
\begin{equation}
    u(t,x) = u_0 e^{i(k \rho_\mathrm{p}\varphi - \omega t)}\cos\bqty{\frac{k_+ + k_-}{2}\rho_\mathrm{p}\varphi - \frac{\omega_+ - \omega_-}{2}t},
\end{equation}
with 
\begin{equation}
    k = \frac{k_+ - k_-}{2}\quad\text{and}\quad\omega = \frac{\omega_+ + \omega_-}{2}.
\end{equation}
This way, we are able to recover the famous result of de Broglie that, through the holonomic constraint (\ref{eq:holonomic}), the frequencies $\Omega_\mathrm{p}$ of the particle appearing in the oscillation $z(\tau)$ and $\omega$ of the field must match according to the relation
\begin{equation}
    \omega = \Omega_\mathrm{p}.
\end{equation}
On top of that, we see that the group velocity for the field must also match the particle velocity
\begin{equation}
    v_g = v_\mathrm{p}.
    \label{eq:velocities}
\end{equation}
More specifically, the relation between $k_\pm$ and $\omega_\pm$
\begin{equation}
    k_\pm = \omega_\pm + \varepsilon_\pm
\end{equation}
lets us write the field
\begin{equation}
    u = u_0 e^{i(k \rho_\mathrm{p}\varphi - \omega t)}\cos\bqty{(\omega + \varepsilon)\rho_\mathrm{p}\varphi - \frac{k - \eta}{\omega + \varepsilon}t},
\end{equation}
with 
\begin{equation}
    \varepsilon = \frac{\varepsilon_+ + \varepsilon_-}{2}\quad\text{and}\quad \eta = \frac{\varepsilon_+ - \varepsilon_-}{2},
\end{equation}
so that equation (\ref{eq:velocities}) gives us
\begin{equation}
    v_g = \frac{k - \eta}{\omega + \varepsilon}\equiv \frac{P_\mathrm{p} - e A_\varphi}{E_\mathrm{p}} = v_\mathrm{p}
    \label{eq:equiv_vel}
\end{equation}
from which we immediately see that $\varepsilon = 0$ and $\eta = e A_\varphi$. This equivalence also obviously implies the relation $(\hbar\omega, \hbar\vb{k}) = (E_\mathrm{p}, \vb{P}_\mathrm{p})$ where we reintroduced the constant $\hbar$ explicitely, meaning that we recover the famous de Broglie hypothesis.
The relation (\ref{eq:equiv_vel}) is fundamental since it justifies the guidance of the particle by the wave.    
Looking at the field, we have in the end two components that we can separate and try to interpret in different ways :
\begin{itemize}
    \item[\textit{i)}] a phase wave following the vertical dynamics of the particle, i.e. its oscillation with frequency $\Omega_\mathrm{p}$, displaying the properties of a monochromatic plane wave and able to introduce some phenomena like quantization rules under specific boundary conditions.
    \item[\textit{ii)}] a group wave following the horizontal dynamics of the particle, i.e. its uniform motion at velocity $v_\mathrm{p}$, displaying the properties of a localized energy packet.
\end{itemize}
This is of course a useful description to understand qualitatively what happens along the particle's trajectory, as de Broglie did, but we will need a more complete and robust mathematical description in order to obtain the field at all space-time points $x$  and derive from it the specific values of its parameters, such as its amplitude and frequencies. Before we tackle the complex field equations however, we will benefit from doing the same for the particle's dynamics, which will be the topic of the next section.

\section{The particle dynamics}

\begin{figure}
    \includegraphics[width=.5\linewidth]{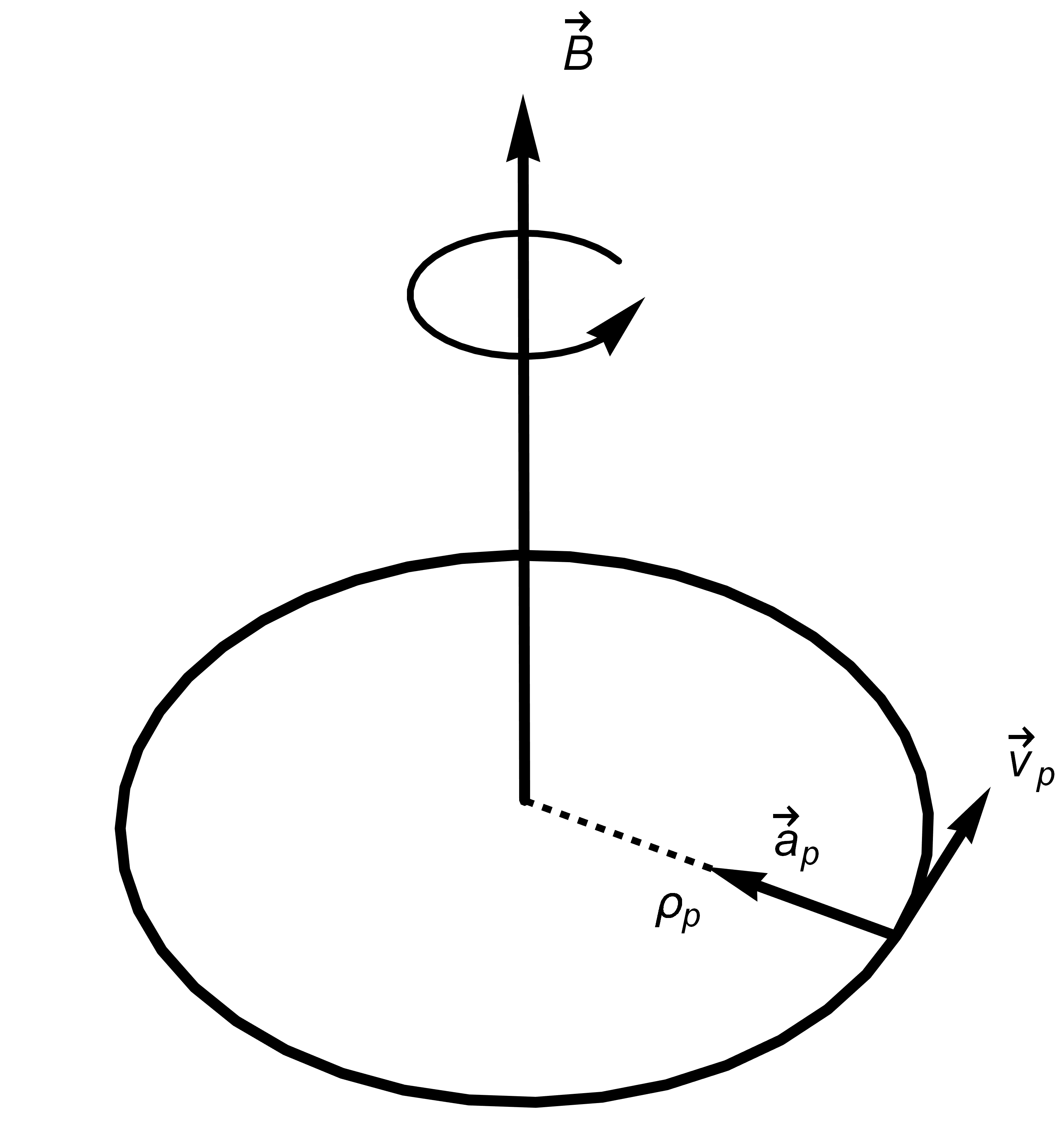}
    \caption{Representation of the particle's motion.}
    \label{fig:figure1}
\end{figure}

Assuming a circular motion in the $(x,y)$ plane with acceleration $\vb{a}_\mathrm{p} = \dv[2]{\vb{x}_\mathrm{p}}{t} = -\frac{v_\mathrm{p}^2}{\rho}\vu{e}_\rho$ as shown on Fig. \ref{fig:figure1}, let us write equation (\ref{eq:motion_x}) again with the specific form of the four-potential $A^\mu = (0, \half B \rho \vu{e}_\varphi)$, i.e., corresponding to a uniform magnetic field $\vb{B} = B\vu{e}_z$ along $z$. The relativistic dynamical equation for the particle reads 
\begin{equation}
    -m_\mathrm{eff.}\gamma\frac{v_\mathrm{p}^2}{\rho} = e v_\mathrm{p} B
    \label{eq:motion_p}
\end{equation} where we used the form of the Lorentz force $e\mathbf{v}_\mathrm{p}\times \mathbf{B}=ev_\mathrm{p} B\vu{e}_\rho$ (note that the electric charge is here negative) and $\gamma=(1-v_\mathrm{p}^2)^{-\frac{1}{2}}$. 
We also define a Bohr-Sommerfeld integral of motion linked to the  angular momentum: 
\begin{equation}
    \oint P_\mathrm{p}\dd{l} = 2\pi \rho\pqty{\gamma m_\mathrm{eff.}v_\mathrm{p} + \half e B \rho} = 2\pi n,
    \label{eq:quant_p}
\end{equation}
with $P_\mathrm{p}$ the linear momentum of the particle, and where $n$ is for now only a quantity translating the conservation of the angular momentum rather than its quantization, meaning that we only assume that it is a real constant. The quantization will of course appear later as we reintroduce the field, but is not necessary at that stage.

From these two equations we derive the velocity
\begin{equation}
    v_n = \frac{1}{\sqrt{1 + \frac{m_\mathrm{eff.}}{2n \omega_{L}}}},
\end{equation}
the radius
\begin{equation}
    \rho_n = \sqrt{\frac{2n}{m_\mathrm{eff.}\omega_{L}}},
\end{equation}
and finally the energy
\begin{equation}
    E_n = m_\mathrm{eff.}\sqrt{1 + 2\frac{n\omega_{L}}{m_\mathrm{eff.}}},
\end{equation}
where we labeled each of the quantites with the index $n$ to indicate that they correspond to a specific orbit of the particle, and where we introduced a frequency
\begin{equation}
    \omega_{L} = -\frac{e B}{m_{\mathrm{eff.}}}
\end{equation}
which we will identify as the usual Landau frequency (we stress that since $e<0$ we have indeed $\omega_L>0$).

We see that there is a dependency on $\omega_L$ in each of the quantities, so that in particular we get for the energy $E_n$ in the non-relativistic limit
\begin{equation}
    E_n \approx m_\mathrm{eff.} + n\omega_{L}.
\end{equation}
This means that as we increase or decrease the intensity of the magnetic field $B$, the energy will vary linearly with it as is expected in these configurations \cite{Landau}. 

\section{Solutions of the field equation}

 In order to derive the solutions for the complex massive scalar field $u$, we start from the Klein-Gordon equation in cylindrical coordinates with covariant derivatives $D^\mu := [\partial_t,-\grad + \half i e B \rho \vu{e}_\varphi]$
\begin{equation}
	-\partial_t^2 u + \frac{1}{\rho}\partial_\rho\pqty{\rho\partial_\rho u} + \frac{1}{\rho^2}\partial_\varphi^2 u + \partial_z^2 u - i e B\partial_\varphi u - \frac{1}{4}e^2 B^2 \rho^2u = \omega_0^2 u,
\end{equation}
giving us a general solution with separate variables
\begin{equation}
    u(t,\rho,\varphi,z) = R(\rho)e^{i(k_z z + m\varphi - \omega t)},\quad m\in \mathbb{Z}.
\end{equation}
Introducing a new variable $\xi = -\half e B \rho^2$, we can extract the equation on $R(\xi)$
\begin{equation}
    \xi R^{\prime\prime}(\xi) + R^\prime(\xi) + R(\xi)\pqty{\beta - \frac{\xi}{4} - \frac{m^2}{4\xi}} = 0
\end{equation}
with $\beta = -\frac{\omega^2 - \omega_0^2 - k_z^2}{2 e B} - \frac{m}{2}$ a constant. This equation admits solutions of the form
\begin{equation}
    R(\xi) = C e^{-\frac{\xi}{2}}\xi^{\frac{\abs*{m}}{2}}{}_1F_1\pqty{-\pqty{\beta - \frac{\abs*{m}+1}{2}},\abs*{m}+1,\xi}
\end{equation}
with $C$ a normalization constant and ${}_1F_1(a,b,z)$ the confluent hypergeometric function solution of 
\begin{equation}
    z f^{\prime\prime}(z) + (b - z)f^\prime(z) + a f(z) = 0.
\end{equation}
Alternatively, this solution is often written in terms of the equivalent Whittaker function $\mathcal{M}_{\kappa,\mu}(z)$, giving us the radial part
\begin{equation}
    R(\xi) = C \sqrt{\xi}\mathcal{M}_{\beta,\frac{\vqty{m}}{2}}(\xi).
\end{equation}

As usual we will get the total field by superposing two counterpropagating solutions
\begin{equation}
    u_\pm(t,\rho,\varphi,z) = C_\pm R\pm(\xi)e^{i(\pm m_\pm + k_{z\pm}z - \omega_\pm t)}.
\end{equation}

If we impose that the radial function be finite everywhere, the quantity $\beta - (\abs{m}+1)/2$ must equal an integer $n_\rho \in \mathbb{N}$ and, knowing the expression of $\beta$, we deduce the value of the energy $\omega$
\begin{equation}
    \omega = \sqrt{\omega_0^2 -2e B \pqty{n_\rho + \frac{\abs{m}+m+1}{2}} + k_z^2}.
\end{equation}
We will now restrict ourselves to cases where $k_z = 0$, meaning to planar motions rather than helicoidal ones. Knowing also that we defined $m_+$ and $m_-$ to be respectively positive and negative integers, the total energy of the field may be defined with the energies of the two counterpropagating waves $\omega_\pm$ as
\begin{equation}
    \omega = \frac{\omega_+ + \omega_-}{2} = \half \omega_0\bqty{\sqrt{1 - 2\frac{eB}{\omega_0^2}\pqty{n_{\rho+} + m_+ + \half}} + \sqrt{1 - 2\frac{eB}{\omega_0^2}\pqty{n_{\rho-} + \half}}}
\end{equation}
which in the limit $-eB/\omega_0^2 \ll 1$ approximates to 
\begin{equation}
    \omega \approx \omega_0 - \frac{eB}{\omega_0}\frac{n_{\rho+}+n_{\rho-} + m_+ + 1}{2}.
\end{equation}
This is equivalent to Landau's formula for the energy $\omega = \omega_0 + n \omega_{L}$ with the quantum number $n$
\begin{equation}
    n := \frac{m_+ - m_-}{2} = \frac{n_{\rho+} + n_{\rho-} + m_+ + 1}{2}\label{truc}
\end{equation}
on the condition that the mass of the field and the particle are equal
\begin{equation}
    \omega_0 = m_{\mathrm{eff.}}.
\end{equation}
This last point is especially interesting since it highlights the very constrained nature of this model. Indeed, to correctly recover the weak field limit, it is for one necessary to have a massive scalar field with $\omega_0 \neq 0$, but also that this mass be equal to the mass of the particle, and finally that the  particle quantum number $n$ of Section II matches $\frac{m_+ - m_-}{2}$ of Eq.~\ref{truc}. This does not mean that a massless field is impossible to work with, as was demonstrated when we introduced a central Coulomb potential in a previous article, but it emphasizes once again the dual nature of our two objects -- that is, the field and particle -- which constrain one another and their respective dynamics.
\section{Conclusion and perspectives}
We developed a quantum mechanical analog of the Landau quantization of the cyclotron orbits of  a charged particle in a uniform magnetic field \cite{Landau}. The model is based on a version of the double solution theory  that we already presented in \cite{Jamet1,Jamet2,Jamet3}. Our approach starts with a relativistically covariant theory in which a particle is interacting with a wave $u(x)$. The model therefore circumvents the usual objection made against Bohmian mechanics concerning action/reaction.   Here however, we have a so called transparency regime which is a particular case of a much more complicated dynamics. In this regime the wave and the particle are not interacting.   It could be interesting to consider the more general (chaotic) regime perhaps associated with transitions between orbits.   We mention that the model is rather peculiar since it relies on a extended wave guiding the particle but the physical reason for this wave is not explained in our model (the causality and the boundary conditions are imposed by hand).  We remind that in the double solution theory of de Broglie \cite{Broglie3} the idea was to find a localized wave  (like a soliton) for defining the particle guided by a base extended wave.  But here, we have still a kind of dualistic dynamics with a particle and an extended wave. In a  subsequent work we are going to show how to modify the present model in order to agree with the core idea of the double solution theory by introducing a localized wave.

\end{document}